\documentclass[iop,apj,tighten]{emulateapj}
\usepackage{graphicx}
\usepackage{amsmath}
\usepackage{amsfonts}
\usepackage{verbatim}
\usepackage[dvipsnames]{xcolor}
\usepackage{hyperref}
\makeatletter
\newcommand{\printfnsymbol}[1]{%
  \textsuperscript{\@fnsymbol{#1}}%
}
\makeatother

\shorttitle{HIGAN: Cosmic neutral hydrogen with GANs}
\shortauthors{Zamudio-Fernandez et al.}

\begin{document}

\title{HIGAN: Cosmic Neutral Hydrogen with Generative Adversarial Networks}

\author{Juan Zamudio-Fernandez\altaffilmark{$^{1}$$^{\dagger}$$^{*}$}}
\author{Atakan Okan\altaffilmark{$^{1}$$^{\dagger}$$^{\star}$}}
\author{Francisco Villaescusa-Navarro\altaffilmark{2} }
\author{Seda Bilaloglu\altaffilmark{$^{1}$}}
\author{Asena Derin Cengiz\altaffilmark{$^{1}$}}
\author{Siyu He\altaffilmark{$^{3}$ $^{2}$}}
\author{Laurence Perreault Levasseur\altaffilmark{$^{2}$}}
\author{Shirley Ho\altaffilmark{$^{2}$$^{3}$$^{4}$} }

\affil{$^{1}${Center for Data Science, New York University, USA }}
\affil{$^{2}${Center for Computational Astrophysics, Flatiron Institute, New York, USA }}
\affil{$^{3}${Department of Physics, Carnegie Mellon University, USA}}
\affil{$^{4}${Department of Astrophysical Sciences, Princeton
University, USA }} 

\altaffiltext{$^{\dagger}$}{Equal contribution}
\altaffiltext{$^{*}$}{jjz289@nyu.edu}
\altaffiltext{$^{\star}$}{ao1512@nyu.edu}

\begin{abstract}
One of the most promising ways to observe the Universe is by detecting the 21cm emission from cosmic neutral hydrogen (HI) through radio-telescopes. Those observations can shed light on fundamental astrophysical questions only if accurate theoretical predictions are available. In order to maximize the scientific return of these surveys, those predictions need to include different observables and be precise on non-linear scales. Currently, one of the best ways to achieve this is via cosmological hydrodynamic simulations; however, the computational cost of these simulations is high -- tens of millions of CPU hours. In this work, we use Wasserstein Generative Adversarial Networks (WGANs) to generate new high-resolution ($35~h^{-1}{\rm kpc}$) 3D realizations of cosmic HI at $z=5$. We do so by sampling from a 100-dimension manifold that the generator learns to map to samples that match closely the fully non-linear abundance and clustering of cosmic HI from the state-of-the-art simulation IllustrisTNG. We show that different statistical properties of the produced samples -- 1D PDF, power spectrum, bispectrum, and void size function -- match very well those of IllustrisTNG, and outperform state-of-the-art models such as Halo Occupation Distributions (HODs). Our WGAN samples reproduce the abundance of HI across 9 orders of magnitude, from the Ly$\alpha$ forest to Damped Lyman Absorbers. WGAN can produce new samples five orders of magnitude faster than hydrodynamic simulations.
\keywords{astrophysics, \and neutral hydrogen, \and generative adversarial networks}
\end{abstract}

\section{Introduction}
The standard model of cosmology is a theoretical framework that accurately explains a large variety of cosmological observations spanning a wide range of spatial and time scales, from the statistical properties of fluctuations in the cosmic microwave background in the early Universe to the spatial distribution of galaxies today. The parameters of this model represent fundamental quantities like the geometry of the Universe, the amount and nature of dark matter and dark energy, and the sum of the neutrinos masses. In the coming years, powerful cosmological surveys -- Giant Meterwave Radio
Telescope (GMRT)\footnote{\url{http://gmrt.ncra.tifr.res.in/}}, the Ooty
Radio Telescope (ORT)\footnote{\url{http://rac.ncra.tifr.res.in/}}, the
Canadian Hydrogen Intensity Mapping Experiment
(CHIME)\footnote{\url{http://chime.phas.ubc.ca/}}, the Five hundred meter
Aperture Spherical Telescope
(FAST)\footnote{\url{http://fast.bao.ac.cn/en/}},
Tianlai\footnote{\url{http://tianlai.bao.ac.cn}}, BINGO (Baryon acoustic
oscillations In Neutral Gas
Observations)\footnote{\url{http://www.jb.man.ac.uk/research/BINGO/}}, ASKAP
(The Australian Square Kilometer Array
Pathfinder)\footnote{\url{http://www.atnf.csiro.au/projects/askap/index.html}},
MeerKAT (The South African Square Kilometer Array
Pathfinder)\footnote{\url{http://www.ska.ac.za/meerkat/}}, HIRAX (The
Hydrogen Intensity and Real-time Analysis
eXperiment)\footnote{\url{https://www.acru.ukzn.ac.za/~hirax/}} and the
SKA (The Square Kilometer
Array)\footnote{\url{https://www.skatelescope.org/}} -- will be collecting data with the aim of constraining the value of these parameters with the highest accuracy possible, in order to improve our understanding of the fundamental physical laws and constituents of the Universe. 

The strategy of some of these billion-dollar-worth surveys is to use radio-telescopes to detect the 21cm emission from cosmic neutral hydrogen (hereafter HI) using the so-called intensity mapping technique \citep[e.g., ][]{Bharadwaj_2001A,Bharadwaj_2001B,Battye:2004re,McQuinn_2006,Chang_2008,Loeb_Wyithe_2008,Bull2015,Santos_2015,Villaescusa-Navarro_2015a}: perform low-angular resolution observations to measure the total 21cm intensity from the HI of many unresolved galaxies and blobs. The abundance and spatial distribution of HI, directly observable with these surveys, contain a large amount of information on fundamental physical quantities. In order to extract it, one must compare the data from these missions against accurate and precise theory predictions.

On large cosmological scales, where spatial fluctuations are small and linear, accurate theoretical predictions can be obtained analytically from first principles of physics. However, in order to extract the maximum information from cosmological surveys (i.e. to constrain the value of the cosmological parameters with the highest accuracy), we need to go into the mildly to fully non-linear regime (i.e. on smaller scales), where analytical predictions from first principles fail. 

Given the very complicated physics that dominates on those small scales, one of the best ways to obtain such predictions is to perform cosmological hydrodynamic simulations. These incorporate the main ingredients that, we believe, control the formation and evolution of galaxies and distribution of cosmic HI, which are the ultimate observables of cosmological surveys. The main challenge of this approach lies in the fact that these simulations are very computationally expensive; e.g. the current state-of-the-art hydrodynamic simulations from \textsc{IllustrisTNG} required more than $\simeq150$ million CPU hours \citep[][]{Pillepich:2017jle,10.1093/mnras/stw2944,10.1093/mnras/stx3112,10.1093/mnras/sty618,10.1093/mnras/stx3040,10.1093/mnras/stx3304,10.1093/mnras/sty2206}. Even with such significant computational investments, the mass and spatial requirements of these simulations only allow running them on relatively small cosmological volumes, limiting their usage on data analysis of large volumes of the Universe. We note that a different direction to make predictions is to combine gravity only simulations with either analytic approaches or results from hydrodynamic simulations \citep{Villaescusa-Navarro_2015a,Villaescusa_Navarro_2018,Modi_2019}. The drawback of this approach is that it relies on relatively simple parametrizations of the HI-halo relation and the HI density profile. Besides, in these models, the effect of the environment is neglected; \cite{Villaescusa_Navarro_2018} showed that environment can play a very important role on setting the clustering of HI.

In order to use observational data to discriminate between different fundamental physics models, fast, accurate, and precise theory predictions, for different observables, ranging from the linear to the fully non-linear regime are needed. In this work, we investigate the use of Generative Adversarial Networks (GANs) to quickly generate high-resolution 3D distributions of cosmic neutral hydrogen with the same statistical properties as the computationally expensive state-of-the-art simulation IllustrisTNG. This constitutes a significant first step towards an efficient generation of accurate theoretical predictions of cosmic neutral hydrogen from large to small scales, needed to maximize the scientific return of upcoming 21cm surveys. We achieve a factor of $\sim10^5$ improvement in terms of computational efficiency; considering 18 million CPU hours for IllustrisTNG100-1 (the simulation we have trained on) and just $<2$ seconds for the production of new samples for using our WGAN model. The $10^5$ factor assumes than a new IllustrisTNG100-1 simulation is produced and all its sub-samples are compared against those of the GAN.

This paper is organized as follows. We briefly discuss some of the works related to the present paper in section \ref{sec:related_work}. In section \ref{section:data}, we describe the data we have used to train our generative models.  We then present the generative models we use and the benchmark model in sections \ref{sec:models} and \ref{sec:benchmark}, respectively. In section \ref{sec:results}, we present the main results of our work and we outline the conclusions in section \ref{sec:conclusions}. 

\subsubsection{Code} Our code and the data we have used are publicly available at \url{https://github.com/jjzamudio/HIGAN}.

\section{Background and Related Work}
\label{sec:related_work}
\subsection{Background}
The standard model of cosmology is a theoretical framework that describes a large variety of cosmological observations. This model of the Universe only requires a few parameters (less than 10), and is usually coined the standard $\Lambda$CDM model, where CDM stands for cold dark matter and $\Lambda$ represents the cosmological constant. A common unit of distance in cosmology is the megaparsec/h ( Mpc/h ) used above is time-dependent, where 1 Mpc is equivalent to 3.26 million light years. The expansion of the Universe stretches the wavelengths, or redshifts, the light that is emitted from distant galaxies, with the amount of change in wavelength depending on their distances and the cosmological parameters. Consequently, for a fixed cosmology we can use the directly observed redshift $z$ of galaxies as a proxy for their distance away from us and/or the time at which the light was emitted.

In this work we focus our attention on cosmic HI. A common way to characterize HI absorbers/emitters is by their column density
\begin{equation}
    N_{\rm HI}=\frac{1}{m_H}\int \rho_{\rm HI}(r)dl
\end{equation}
where $m_H$ is the mass of the hydrogen atom, $\rho_{\rm HI}$ is the density of HI in the considered absorber/emitter and the integral is carried out over the entire system along the line-of-sight. Systems with HI column densities below $\simeq10^{17.3}~{\rm cm}^{-2}$ are called the Ly$\alpha$-forest, and physically represent the low-density and highly ionized gas that resides in the intergalactic medium (IGM). Systems with column densities above $10^{20.3}~{\rm cm}^{-2}$ are coined Damped Lyman Absorbers (DLAs). They are self-shielded against external radiation and it is believed that they correspond to extra-galactic regions. Finally, systems with column densities between $\simeq10^{17.3}~{\rm cm}^{-2}$ and $10^{20.3}~{\rm cm}^{-2}$ are called Lyman Limit systems (LLSs) and they reside is a much broader environment between the IGM and the dense galactic regions \citep[][]{Villaescusa-Navarro_2015a,Villaescusa_Navarro_2018}. In this work we will show how WGAN can model HI from the Ly$\alpha$-forest to DLAs, across more than 9 orders of magnitude in HI column density.

\subsection{Related Work}
Recent progress in machine learning has caused a rapid increase in applications of deep learning to cosmology, both in number of instances and in variety of problems addressed. In one of the earliest examples of deep learning application to cosmology, \cite{Ravanbakhsh:2017bbi} used deep 3D convolutional neural networks (CNNs) to predict cosmological parameters from the matter distribution (see \cite{2018arXiv181001483C} for parameter estimation from CMB lensing). Following this work, \cite{Schmelzle:2017vwd} applied a deep CNN (DCNN) to discriminate between cosmological models using weak lensing maps, which represent projected 2D dark matter distributions. These works were extented by \cite{Peel_2018,Merten_2019}, who used CNNs to distinguish between different cosmological models, including modified gravity and massive neutrinos. More recently, \cite{1811.06533} build a DCNN able to predict computationally expensive $N$-body simulations from the linear perturbation theory, enabling, for the first time, the fast approximate simulation of complex 3D physical system using deep learning. In \cite{1902.05965}, the authors built a two-stage CNN to map from the dark matter field of an N-body simulation to the galaxy distribution of the full hydrodynamic simulation Illustris. The authors demonstrated the ability of their network to reproduce many statistical properties of the Illustris galaxy field such as power spectrum, bispectrum and abundance of voids down to very small scales. This suggests that flexible, high-dimensional interpolation allowed by deep neural networks can accurately capture the mapping from dark matter to galaxies. Finally, \cite{2018arXiv181008211L} trained CNNs to extract information from 21cm maps from the epoch of reionization, showing that they can accurately recover the duration of the epoch of reionization.

Regarding unsupervised and generative methods applied to cosmology problems, \cite{Mustafa2017CreatingVU} used GANs to generate weak lensing convergence maps, producing samples that match closely the statistics of the true lensing convergence maps. \cite{10.1093/mnrasl/slx008} used GANs to recover features from artificially degraded galaxy images. \cite{Shirasaki_2018} used conditional adversarial networks to denoise weak lensing maps. \cite{Rodriguez2018} generated 2D mass distributions realizations of the cosmic web using different GANs. \cite{Ravanbakhsh:2016xpe} developed deep conditional generative models to produce higher quality images of galaxies. Finally, \cite{Ramanah:2019cbm} implemented a Wasserstein GAN that successfully learns to map 3D dark matter fields to realistic halo distributions.

\section{Data}
\label{section:data}
 
The simulation used in this work as training data is TNG100-1, produced by the IllustrisTNG collaboration \citep[][]{Pillepich:2017jle}. This simulation represents the current state-of-the-art in terms of cosmological hydrodynamic simulations. It implements a wide range of relevant physical effects, such as radiative cooling, star-formation, metal enrichment, supernova and AGN feedback, and magnetic fields effects. The simulation was initialized at redshift $z=127$, i.e. 12 million years after the Big Bang. Approximately 6 billion dark matter particles and 6 billion gas cells were then evolved for $\sim$ 13.8 billion years, down to the present epoch, in a volume of $(75~h^{-1}{\rm Mpc})^3$ (a cubic volume of $\sim$250 million light years per side). For further details, we refer the reader to \cite{Pillepich:2017jle}.

In this work, we focus our attention to cosmic HI at redshift 5 (1.2 Gyr after the Big Bang), whose abundance and spatial distribution is modelled using the output of IllustrisTNG and the post-processing work of \cite{Villaescusa_Navarro_2018}. This redshift was chosen for several reasons: 1) this epoch in the history of the Universe is a target for multiple upcoming 21cm surveys such as SKA1-LOW and future radio-telescopes \citep[][]{Ansari:2018ury,Obuljen_18}, 2) it is known that the current state-of-the-art methods to model the abundance and distribution of HI at this epoch face several important challenges \citep[][]{Villaescusa_Navarro_2018,Villaescusa-Navarro_2015a}, and 3) the sparsity of the field is smaller at this redshift than at lower redshift, facilitating the training of deep learning models\footnote{We have observed that, at lower redshifts, the training of neural networks is more challenging due to the high sparsity and large variance of the HI field.}.

We use the cloud-in-cell (CIC) interpolation scheme to assign positions and HI masses from voronoi gas cells onto a regular grid with $2048\times2048\times2048$ cells. The value stored in each cell of the grid is thus the total HI mass (or equivalently, the local HI density). This grid size allows us to sample the HI field at a very high spatial resolution: $\simeq35~h^{-1}{\rm kpc}$. We have kept a cubic section with a volume equal to 1/8 times that of the entire cube as test set. For the training data, we randomly sample sub-cubes of 64 $\times$ 64 $\times$ 64 cells, corresponding to a comoving volume of $\sim(2.34~h^{-1}{\rm Mpc})^3$, from the region outside of the test set. We have taken this number of cells for the training sub-cubes as a compromise between 1) generating big cosmological volumes and 2) having enough samples to train the network.

\subsubsection{Data Preprocessing}

The probability distribution function of our data exhibits a very long tail for cells with large HI masses (see table \ref{tab1}). In order to facilitate the training of our model, we perform the following transformation to scale the data to the $[-0.23, 1]$ interval:
\begin{equation}
\label{data-transformation}
\begin{aligned}
\tilde{m}_{\rm HI} = \frac{\log_{10}(m_{\rm HI} + \epsilon)}{ \log_{10}(m_{\rm HI}^{\rm max} + \epsilon)}
\end{aligned}
\end{equation}
where $m_{\rm HI}^{\rm max}$ is the maximum HI mass from all cells and we have set $\epsilon=0.01$.

\begin{table}
\begin{center}
\caption{Statistical properties of our training data}\label{tab1}
\begin{tabular}{|c|c|c|c|c|}
\hline
{\bf Min} & {\bf Max} & {\bf Median} & {\bf Mean} & {\bf SD}\\
\hline 
0 & $4.4\times10^9$ & 13.2 & $1.5\times10^4$ & $9.2\times10^5$\\
\hline
\end{tabular}

\end{center}
\end{table}

\section{Architecture}
\label{sec:models}

In its simplest form, a GAN consists of two networks: a generator and a discriminator \citep[][]{NIPS2014_5423}. The former, referred to as the generator, tries to generate samples similar to the real data, while the latter, referred to as the discriminator or critic, tries to differentiate the generated samples from the real ones. This interaction can be seen as a zero-sum value game where the discriminator maximizes the probability of classifying each sample correctly, while the generator tries to maximize the probability of the discriminator classifying generated samples as real.

In the traditional GAN training, the discriminator is trained to maximize
\begin{equation}
    L(D,g_\theta) = E_{x \sim P_r}[\textrm{log } D(x)] + E_{x \sim P_{\theta}}[\textrm{log }(1-D(x))]
\end{equation}
where $P_r$ represents the distribution of the real data (in our case, the IllustrisTNG simulation), $P_{\theta}$ is the distribution of the parametrized density, and $D(x)$ is the probability that $x$ comes from real data rather than $P_{\theta}$. \cite{DBLP:journals/corr/ArjovskyCB17} showed that, in equilibrium, this setting is equivalent to minimizing the Jensen-Shannon (JS) divergence between the distributions of the real samples and the generated ones. JS is however locally saturated, which leads to vanishing gradients and therefore severely affects the training process. 

To tackle this problem, the authors proposed the Wasserstein GAN. Instead of minimizing the JS divergence, the Wasserstein GAN minimizes the Earth-Mover (EM) distance, which is continuous and differentiable almost everywhere, providing smoother gradients that help training the critic. In this context, the critic, instead of trying to differentiate between real and generated samples, provides an approximation of how far the generated samples are from the real ones. In practice, however, the calculation of the EM distance is highly intractable. Under certain assumptions, the problem can be approximated as training a network parametrized by weights that lie in a compact space by clamping the critic's weights. \cite{DBLP:journals/corr/GulrajaniAADC17} showed that using gradient penalty in the critic's loss function instead of weight clamping is a significantly better approach to ensure that the Lipschitz constraint is met, while also allowing the generator to learn more complex representations. 

In this work, we implement the Wasserstein GAN with gradient penalty, where the inputs to the critic and outputs of the generator are cubes of size $64 \times 64 \times 64$. The critic's loss function is given by: 
\begin{equation} \label{critics-loss}
\begin{split}
L_c =  \mathbb{E}_{x \sim P_r}[f_w(x)] -\mathbb{E}_{z\sim P_z}[f_w(g_\theta(z))]
\\
+ \lambda   \mathbb{E}_{\hat x \sim P_{\hat x} } [(\left\| \nabla _{\hat{x}} f_w (\hat x) \right\|_2 - 1)^2 ]
\end{split}
\end{equation}
where $f_w$ is the critic, $g_\theta$ is the generator, $z$ is a vector of size $100$ sampled from a standard Gaussian distribution, and $\lambda$ is the gradient penalty coefficient which we set equal to 10.  $\hat x$ is an interpolation between the real and generated samples: $\hat{x}= tx +(1-t)y$, where $t \sim U[0,1]$ and $x \sim P_r, y \sim P_z$.

The loss function of the generator is given by: 

\begin{equation}\label{generator-loss-wgan}
\begin{aligned}
L_G =   -\mathbb{E}_{z\sim P_z}[f_w(g_\theta(z))] 
\end{aligned}
\end{equation}

For the networks we use an all-convolutional architecture, similar to the deep convolutional GAN (DCGAN) implementation of \cite{DBLP:journals/corr/RadfordMC15}, which we adapt to three-dimensional data. The generator consists of five 3D deconvolutional (transpose convolution) layers with rectified linear unit (ReLU) activation and batch norm (3D) layers followed by two convolutional layers, with filter sizes 3 and 2, respectively. The final activation function is tanh. The number of channels in the first layer is 1024, halving in size until the last layer which receives 128 channels and outputs 1. The critic has seven convolutional layers with leaky ReLU activations. The number of channels is symmetric to the generator. 

During our experiments, we found that adding the extra convolutional layers without changing the feature map sizes improved significantly the quality of the generated samples. This is presumably because, theoretically, it is known that the problem at hand is highly non-linear and the HI distribution at cosmological scales cannot be easily parametrized. Thus, the addition of layers allows the injection of more non-linear activations to the model, enabling the network to learn more complex representations of the IllustrisTNG data.

We used the Adam optimization algorithm with learning rate of 0.0005 and betas of 0.5 and 0.9. We did not find a need for extensive hyperparameter optimization. We used a batch size of size 16. We trained all of our models using New York University's High Performance Cluster using NVIDIA Tesla P-100 GPUs. We trained for approximately 400 hours and 150,000 generator iterations.

We also implemented the Maximum-Mean-Discrepancy GAN (MMD-GAN) \citep[][]{DBLP:journals/corr/LiCCYP17} in addition to the WGAN for the generation of two dimensional samples (obtained by projecting the cubes along one axis). The choice to implement the WGAN for the generation of 3D data was mainly driven by the results obtained at this stage; our 2D WGAN samples achieved better results, qualitatively and in terms of the 1D-PDF, with significantly less hyperparameter optimization and training time than the MMD-GAN samples. See the Appendix for more details.

\section{Benchmark model: Halo Occupation Distribution}
\label{sec:benchmark}

\cite{Villaescusa_Navarro_2018} showed that, in the post-reionization era ($z<6$), most of the cosmic HI in the Universe resides within dark matter halos. This is the basis principle underlying the state-of-the-art framework developed in \cite{Villaescusa_Navarro_2018} to model the abundance and spatial distribution of HI. This model, which we briefly describe here, is the one used in this work as a benchmark model to judge the performances of our generative model. In what follows, we refer to this method as the Halo Occupation Distribution (HOD) model. We refer the reader to other works for similar approaches \citep[][]{Villaescusa-Navarro_2015a,Castorina_2017,Barnes_2014}.

The first step of the HOD model requires running a gravity-only simulation (e.g. an N-body simulation) and identify the positions and masses of their halos. Next, the total HI mass residing within a given halo is calculated as 
\begin{equation}
M_{\rm HI}(M,z)=M_0\left(\frac{M}{M_{\rm min}}\right)^\alpha \exp(-(M_{\rm min}/M)^{0.35})
\label{eq:M_HI}
\end{equation}
where $M$ is the halo mass and $M_0$, $M_{\rm min}$, and $\alpha$ are free parameters whose values at $z=5$ are\footnote{Those values were calibrated using the TNG100-1 simulation.}: $M_0=1.9\times10^9~h^{-1}M_\odot$, $M_{\rm min}=2\times10^{10}~h^{-1}M_\odot$, and $\alpha=0.74$ \citep[][]{Villaescusa_Navarro_2018}. Finally, the total HI mass is distributed spatially inside the halos. For this procedure, we follow \cite{Villaescusa_Navarro_2018} and assume a HI density profile given by
\begin{equation}
\label{eq:rho_HI}
\rho_{\rm HI}(r)  = \frac{\rho_0}{r^{\alpha_*}}\exp(-r_0/r)
\end{equation}
For simplicity, we consider that the HI mass within 1 kpc/h of the halo center is negligible in all cases, and take $\alpha_*=3$ for all halo masses. This simplification allow us to randomly sample the above distribution is a significantly more efficient way.

More specifically, the procedure we use to generate 3D HI mock distributions is as follows. We first randomly select a sub-cube of TNG100-1-Dark, which is a dark-matter-only N-body simulation with the same mass, spatial resolution, and mode phases as the full hydrodynamical TNG100-1. The volume of this sub-cube is chosen to be equal to those of the training set. We then identify all halos that lie inside that region by finding their positions and masses. For each halo, we then compute its total HI mass using Eq. \ref{eq:M_HI}. We then distribute the total HI mass inside the halo using 1000 particles; the HI mass of each of these particles is $M_{\rm HI}(M,z)/1000$. For each particle, its radius is drawn from the HI density distribution of Eq. \ref{eq:rho_HI} while its direction is taken randomly within the sphere. We repeat this procedure for all halos that lie within the considered sub-cube. Finally, we construct a $64\times64\times64$ grid from the positions and HI masses of all the particle tracers, using the CIC interpolation scheme. This HI grid can be directly compared with either our test set or the WGAN outputs. More details on the HOD model and its accuracy can be found in \cite{Villaescusa_Navarro_2018}.

We note that this model, together with similar approaches \citep[e.g., ][]{Villaescusa-Navarro_2015a,Castorina_2017,Barnes_2014}, were developed to model the 21cm cosmic signal. Since most of the HI in the Universe in the post-reionization ($z<6$) resides within DLAs and LLS, these models perform well when describing those systems, but they are expected to perform poorly for low column density systems such as the Ly$\alpha$-forest. It may be possible to improve the performance of these models on such systems by combining their predictions with other methods such as the Fluctuating Gunn Peterson Approximation \citep[][]{FGPA}. This is however beyond the scope of this work.

\section{Results \& Validation Metrics}
\label{sec:results}

In this section, we present the main results of our generative model, and compare it against the test set of IllustrisTNG and the benchmark model. We quantify the agreement between the 3D HI distribution of the different models using four different summary statistics: the 1D PDF, the power spectrum, the bispectrum and the void size function. We emphasize the importance of using multiple statistics to quantify the agreement between the different samples, as each of them will be sensitive to different regions of the underlying field. In all cases, we have taken 1000 sub-cubes from the IllustrisTNG test set, 1000 sub-cubes from the WGAN and 1000 sub-cubes from the HOD model. For each of these models, we have computed the considered summary statistics on each sub-cube, and calculated the mean and standard deviation of the results.

\begin{figure*}
      \includegraphics[width=0.99\textwidth]{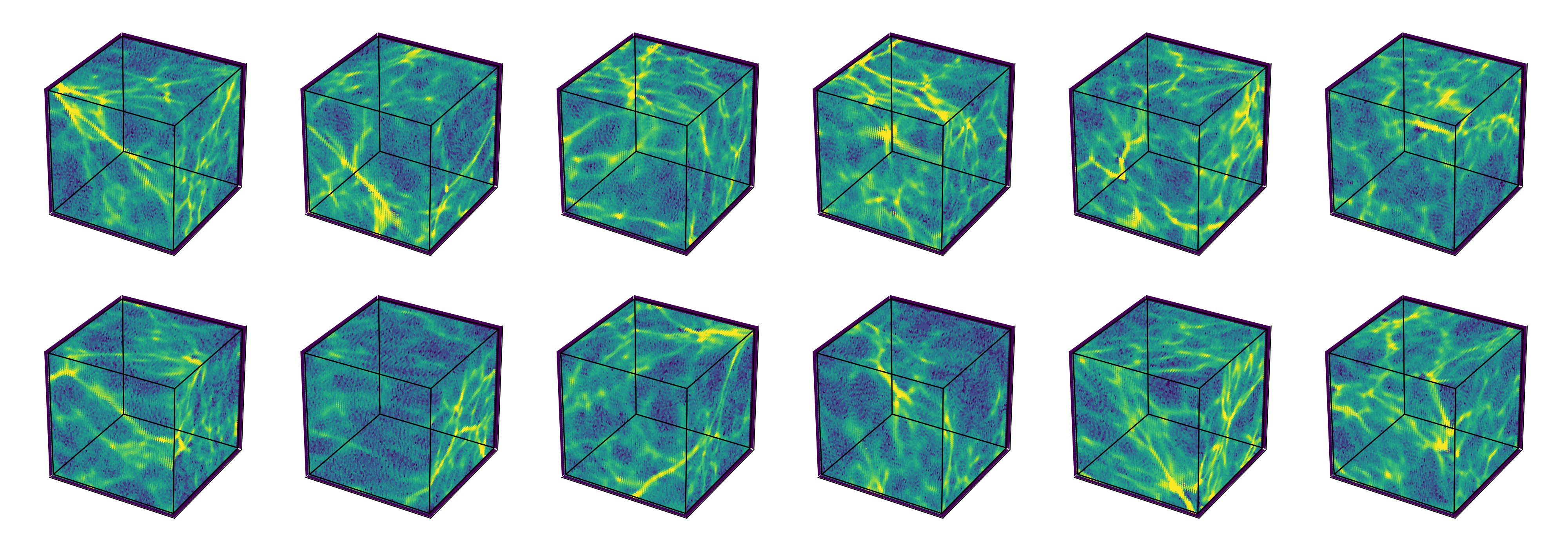}\\
      \vspace{0.5cm}\\
      \includegraphics[width=0.99\textwidth]{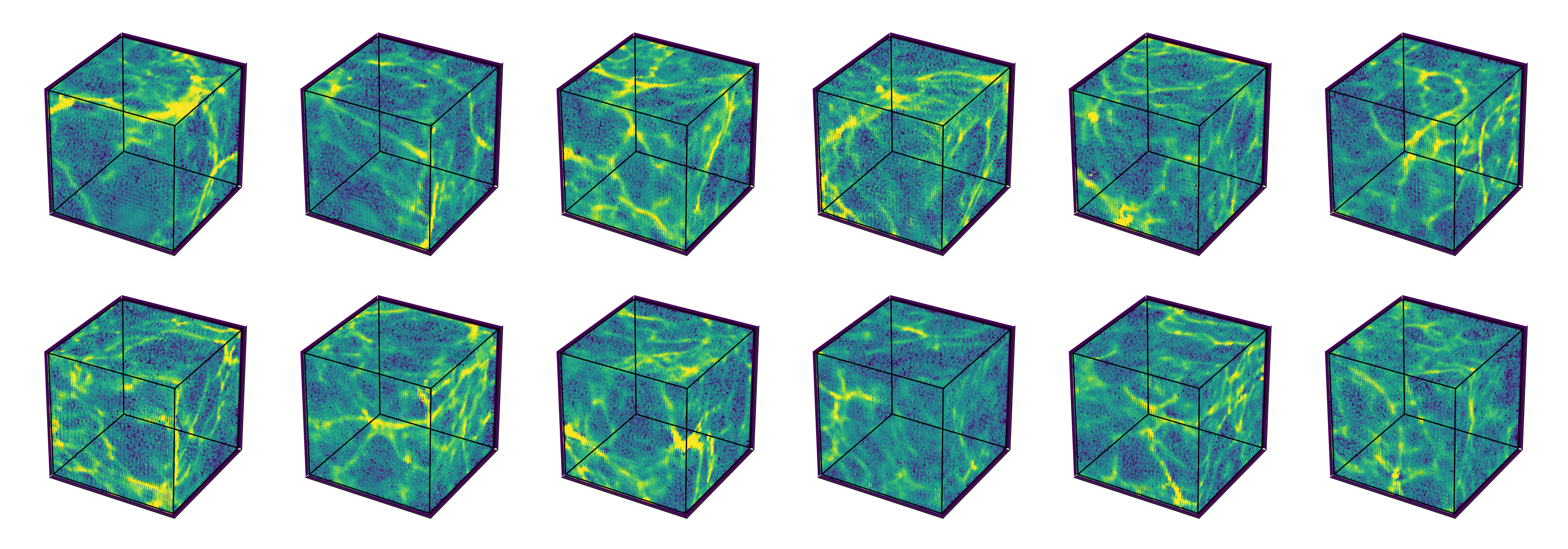}
      \caption{3D HI distributions from IllustrisTNG (top rows) and WGAN (bottom rows). The network successfully produces spatial distributions with all the elements of the HI web: filaments, voids and dense regions.}
      \label{fig1}
\end{figure*}

\begin{figure*}
\includegraphics[width=0.99\textwidth]{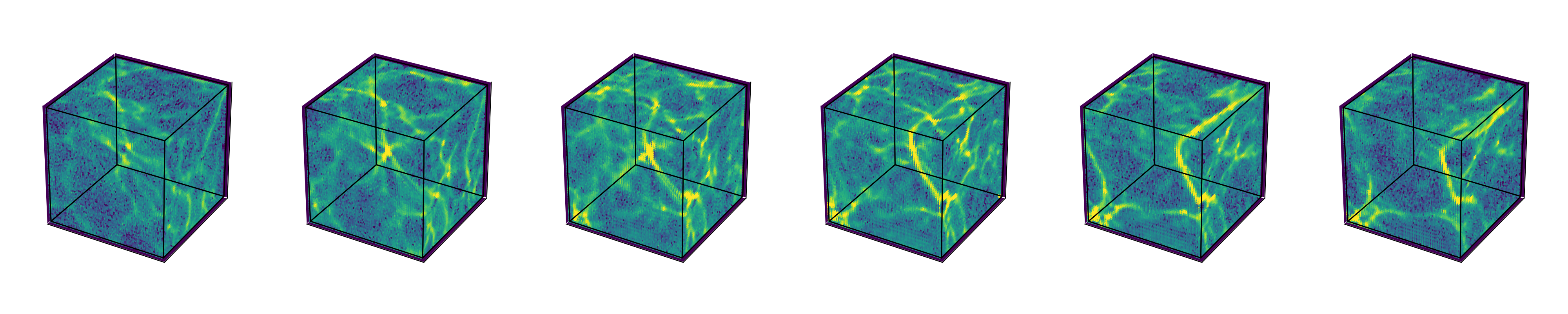}
\caption{WGAN Samples generated by linearly interpolating between two latent space vectors. The generator learns a smooth mapping from the latent space to the output space and is able to generate diverse samples, which demonstrates that it is not simply memorizing the training samples. } \label{fig3}
\end{figure*}

In Fig. \ref{fig1}, we show examples of 3D HI distributions from IllustrisTNG and generated from our trained WGAN. As can be seen, the WGAN is able to generate realizations that visually look very similar the IllustrisTNG samples. Likewise, in Fig. \ref{fig3}, we can see that the generator has learned a smooth mapping from the latent space to the output space.

\subsection{1-D Probability mass function}

\begin{figure}
\begin{center}
    \includegraphics[width=\linewidth]{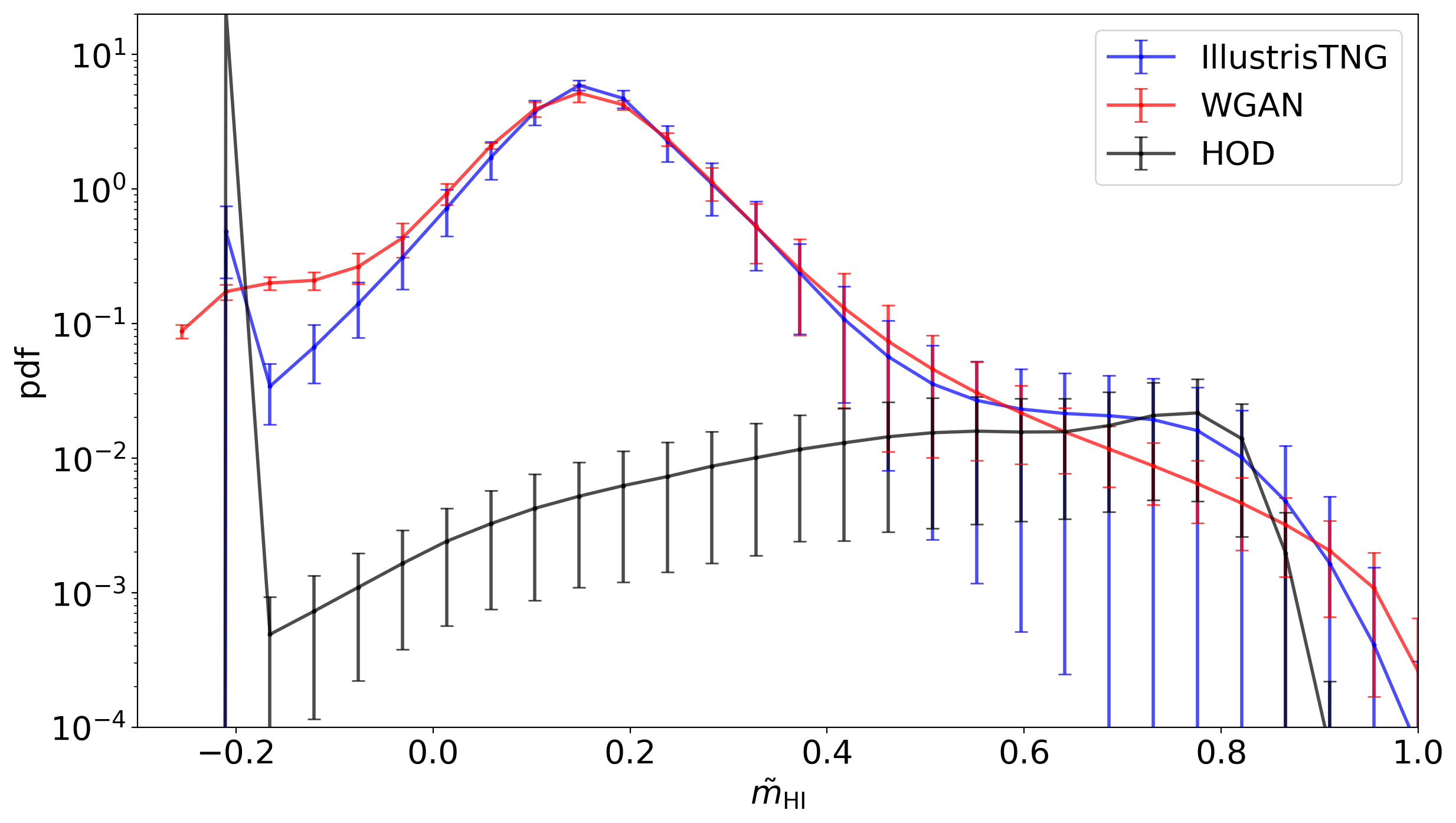}
\end{center}
\caption{We take 1000 sub-cubes from IllustrisTNG (blue), WGAN (red) and HOD (black) and compute the 1D HI PDF for each sub-cube. The plot shows the mean and standard deviation in log scale as a function of HI mass for the three sample sets. The WGAN model reproduces well the HI PDF from IllustrisTNG, while the HOD fails to reproduce the central part of the PDF, which is dominated by low HI density cells corresponding to the Ly$\alpha$-forest. The x axis represents the transformed mass as indicated by Eq\ref{data-transformation} to show the intervals better. -0.23 corresponds to zeroes in real scale while 1.0 corresponds to the maximum value in the IllustrisTNG simulation.}
\label{fig4}
\end{figure}

The first statistic we consider is the 1D PDF. We compute it as the fraction of cells with HI masses lying within a given interval, per unit of HI mass, as a function of HI mass. In our case, the volume of each cell is $\simeq(35~h^{-1}{\rm kpc})^3$, i.e. we are deep into the non-linear regime. The PDF is one of the simplest statistics that inform us on the distribution of HI masses in cells. We show the results of our analysis in Fig. \ref{fig4}. 

We first consider the results of the state-of-the-art HOD model. It is clear from the plot that for low-HI masses cells (corresponding to the Ly$\alpha$-forest), the HOD model is not able to reproduce the HI PDF from IllustrisTNG. As we argued in Sec. \ref{sec:models}, the reason is that these models are designed to model well LLS and DLAs, but not the Ly$\alpha$-forest. We can see that in the high-HI mass tail, the HOD reproduces relatively well the results of IllustrisTNG (considering the large errorbars). 

On the other hand, the WGAN model is able to reproduce the PDF from IllustrisTNG from $\tilde{m}_{\rm HI}\in[0,1]$, i.e. for more than 9 orders of magnitude for HI masses. On the low mass tail, the WGAN model performs slightly worse. In particular, we can see how our model fails at reproducing a peak around -0.23. That peak corresponds to empty cells in IllustrisTNG. We notice however that for 21cm intensity mapping, those cells will have negligible contribution to the observed signal. 

We note that, for some HI masses, the variance of the WGAN samples are smaller than the ones from IllustrisTNG. We believe that this can be addressed by conditioning the GAN on the amplitude of the HI power spectrum in a cell. This is, however, beyond the scope of this work and we plan to explore this direction in a follow-up paper. 

\subsection{Power Spectrum}

\begin{figure}
  \includegraphics[width=\linewidth]{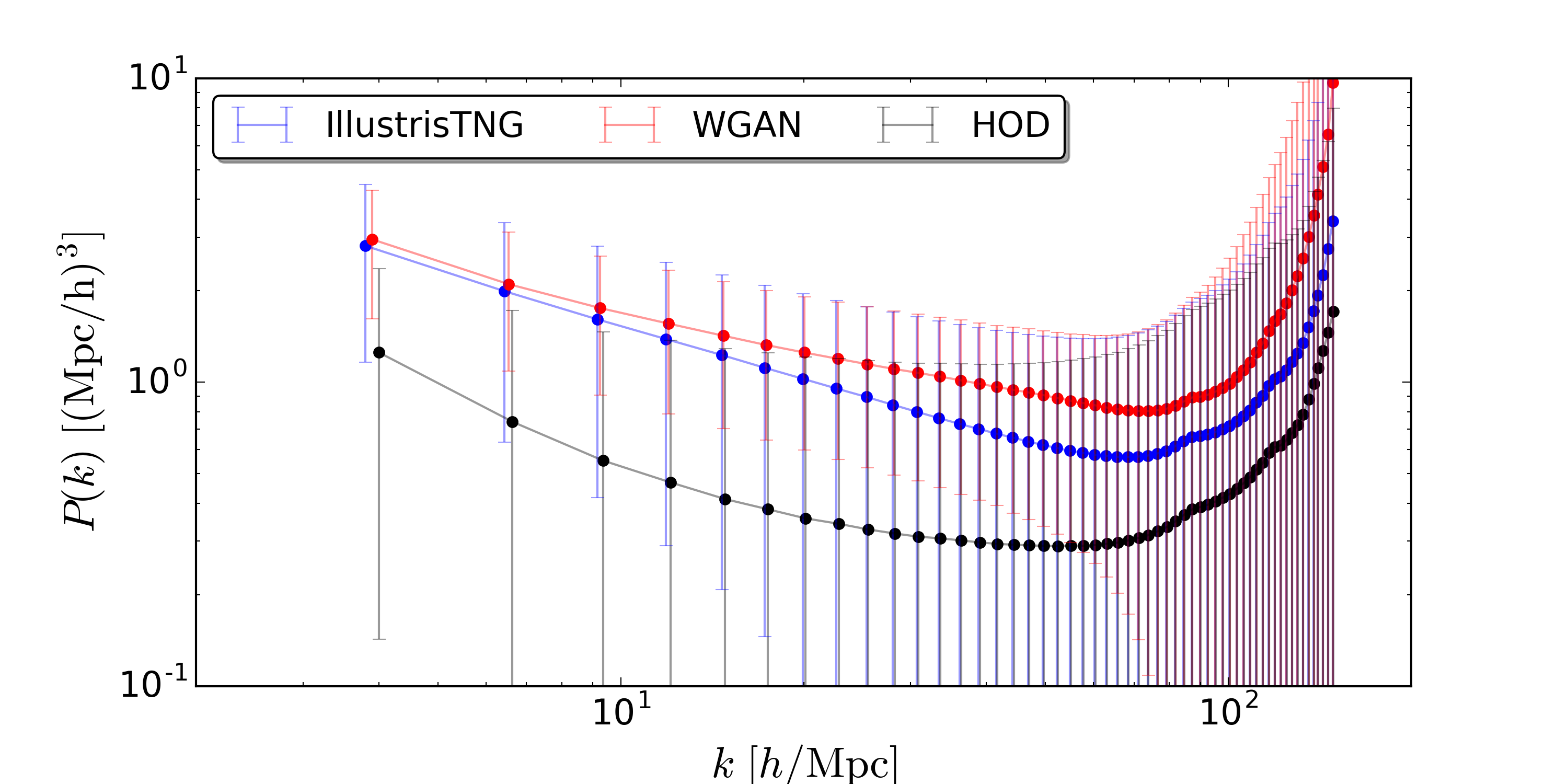}
  \caption{We take 1000 sub-cubes from IllustrisTNG (blue), WGAN (red) and HOD (black) and compute the HI power spectrum for each sub-cube. The plot shows the mean and the standard deviation as a function of wavenumber for the three sets. The WGAN samples have a very similar clustering amplitude and shape as those from IllustrisTNG on large scales, although they slightly deviates on smaller scales. The state-of-the-art HOD model is not able to reproduce the power spectrum from IllustrisTNG on any of the above considered scale. We should note the scales considered here are highly non-linear. The HOD model is expected to perform well on large-scales.}
  \label{fig:ps}
\end{figure}

The second summary statistic we consider is the HI power spectrum. The power spectrum plays a central role in cosmology because, for Gaussian density fields, it is the only summary statistic needed to extract all information contained in the field. The power spectrum is defined as the Fourier transform of the correlation function:
\begin{equation}\label{eq:6}
\begin{aligned}
\xi_{\rm HI}(r)= \langle \delta_{\rm HI}(\vec{r}') \delta_{\rm HI}(\vec{r}'+\vec{r}) \rangle 
\\
P_{\rm HI}(k) = \int d^3 \vec{r} \xi_{\rm HI}(r) e^{i\vec{k}\cdot\vec{r}}
\end{aligned}
\end{equation}
where $\delta_{\rm HI}(\vec{x})=\rho_{\rm HI}(\vec{x})/\bar{\rho_{\rm HI}}-1$. In Fig. \ref{fig:ps} we show the mean HI power spectra and standard deviation from the three sample sets.

We find that the HI distribution from the WGANs samples has, on large scales,  an average power spectrum which is very close in amplitude and in shape to that of the IllustrisTNG samples. We emphasize that on the very small scales we are considering here, matching the clustering amplitude and shape from IllustrisTNG is highly non-trivial. This can be demonstrated by the fact that the state-of-the-art HOD model is not able to reproduce the correct clustering of HI on any of the considered scales. On smaller scales, we find that the power spectrum of the samples produced by our model departs from that of the IllustrisTNG samples, although results are still compatible given the large variance of different realizations. As for the 1D PDF, we find that the variance in the power spectrum of the WGAN samples is smaller than that of both the IllustrisTNG  and HOD model samples.

\subsection{Bispectrum}

The third statistic we study is the HI bispectrum. It is defined as a three-point function in Fourier space:

\begin{equation}\label{eq:7}
B(k_1,k_2,k_3)\delta_K(\vec{k}_1 + \vec{k}_2 + \vec{k}_3)=\langle \delta_{\rm HI}(\vec{k}_1) \delta_{\rm HI}(\vec{k}_2) \delta_{\rm HI}(\vec{k}_3) \rangle
\end{equation}

The bispectrum plays an important role in cosmological analysis, as it carries information pertaining to departure of a field distribution from Gaussianity. In other words, for Gaussian fields, the bispectrum is zero, whereas for non-Gaussian fields the shape and amplitude of the bispectrum contains additional information. For example, in the case of interest here, the bispectrum is more sensitive to the morphology of the HI 3D distribution than the power spectrum. 

\begin{figure}
  \includegraphics[width=\linewidth]{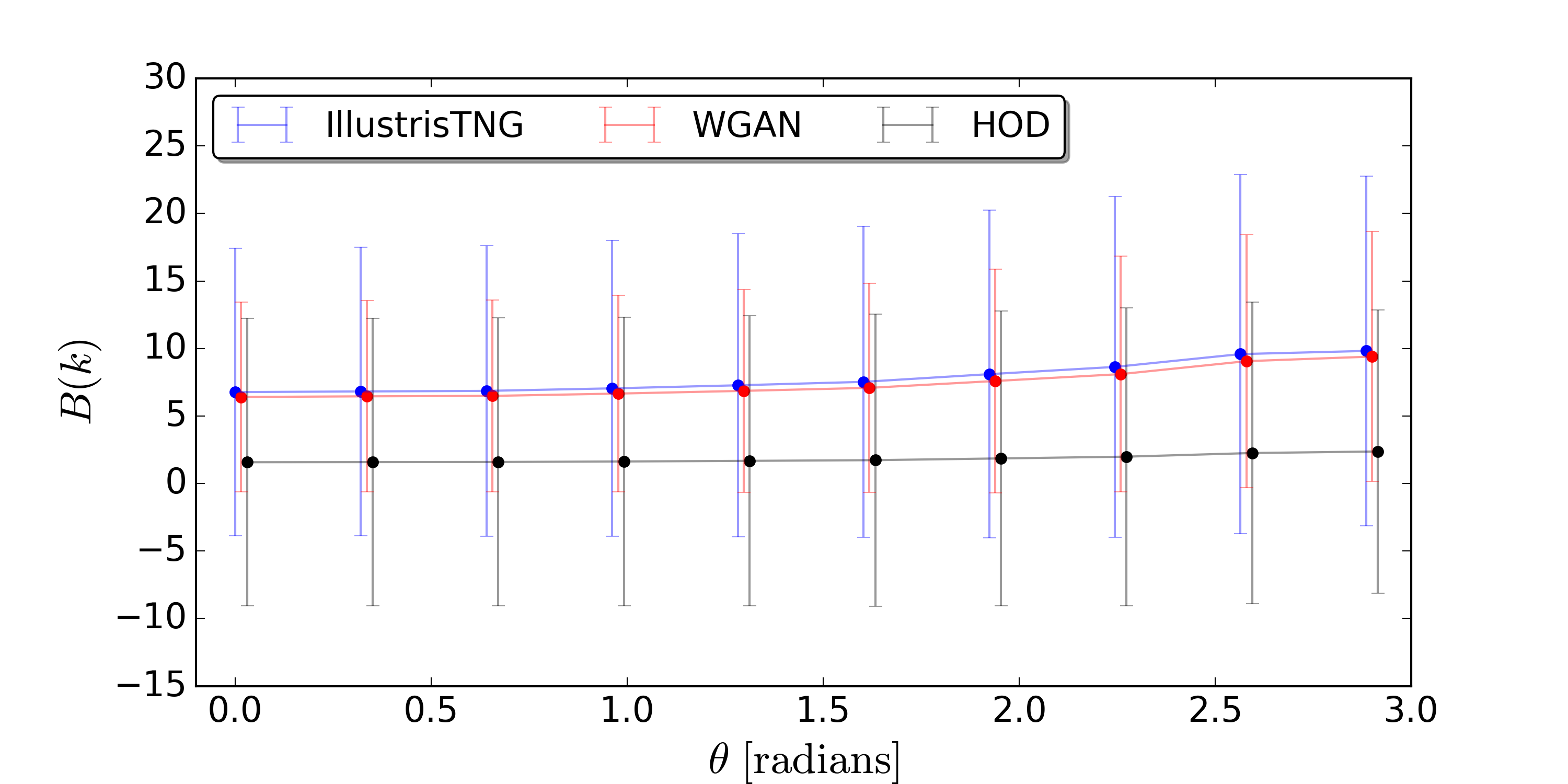}
  \caption{We take 1000 sub-cubes from IllustrisTNG (blue), WGAN (red) and HOD (black) and compute the HI bispectrum for each sub-cube setting $k_1=6~h/{\rm Mpc}$ and $k_2=6.5~h/{\rm Mpc}$. The figure shows the mean and standard deviation as a function of the angle between $k_1$ and $k_2$ for the three sets. We find that the bispectra of the WGAN samples are closer to those of IllustrisTNG than to those of the HOD. We find similar results for other triangle configurations.}
  \label{fig:bs}
\end{figure}

Fig. \ref{fig:bs} shows the mean and standard deviation of the bispectrum from the three different samples. For simplicity, we have selected a value of $k_1=6~h/{\rm Mpc}$ and $k_2=6.5~h/{\rm Mpc}$ and varied the angle $\theta$, i.e. $k_3$, although similar results take place for other configurations.

We find that the sub-cubes produced by the WGAN model have bispectra in closer agreement with the ones from IllustrisTNG than with those from the HOD model, which produces results with a lower amplitude. We notice that, for this statistics, it is also the case that the variance in the WGAN samples is smaller than in the IllustrisTNG and HOD samples.

\subsection{Voids Abundance}

The last statistic we consider is the abundance of cosmic voids in the HI field. By voids we mean connected regions in the HI density field that are underdense with respect to the mean HI density.

The statistics we work with is the void size function, defined as the number density of voids per unit of radius as a function of void radius. This statistic contains complementary information to that of the bispectrum, power spectrum, and 1D PDF. Therefore, it can be seen as an additional probe of the spatial distribution of the cosmic HI field. 

\begin{figure}
  \includegraphics[width=\linewidth]{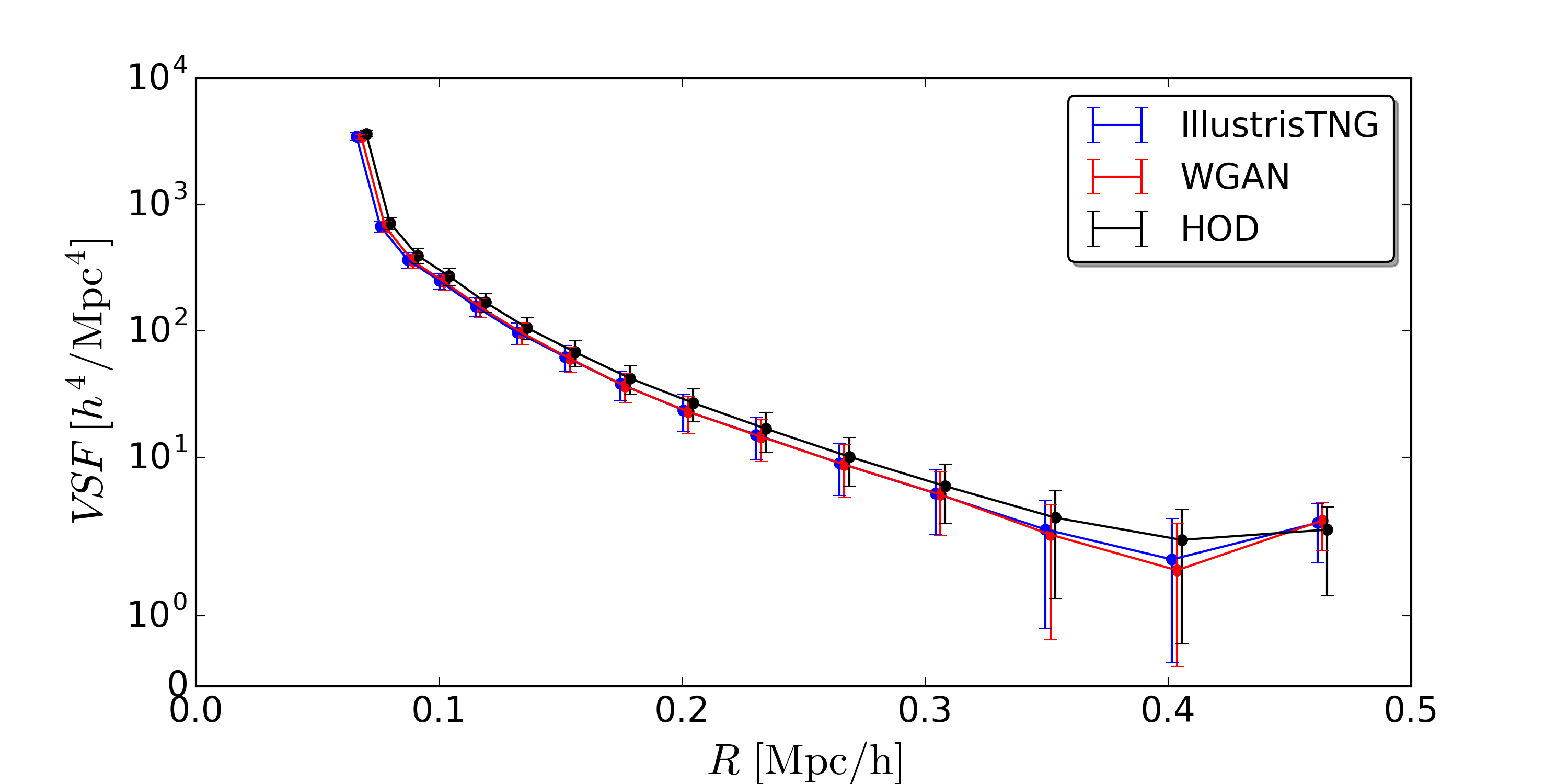}
  \caption{We take 1000 sub-cubes from IllustrisTNG (blue), WGAN (red) and HOD (black) and identify voids (underdense regions) in the HI field for each sub-cube. The plot shows the mean and the standard deviation of the void size function (number density of voids per unit of radius) as a function of radius for the three sets. We find that both the WGAN and HOD models produce fields with the same abundance of underdense regions as IllustrisTNG.}
  \label{fig:void}
\end{figure}

For each of the 1000 sub-cubes of each model, we have identified voids in the neutral hydrogen distribution using the algorithm described in \cite{Banerjee:2016zaa}. We have then computed the mean and the standard deviation of the results. We show the results in Fig. \ref{fig:void}.

We find that both the WGAN and HOD samples have a distribution of HI voids that is in very good agreement with that of the IllustrisTNG samples.

\section{Conclusions and Future Work}
\label{sec:conclusions}

Upcoming cosmological surveys aim at determining the value of the cosmological parameters with the highest accuracy possible, in order to improve our understanding of fundamental physics. To achieve this, the data from these missions need to be compared against accurate and precise theoretical predictions. Since most of the information we seek is encoded in non-linear scales (i.e. small scales), theoretical predictions valid down to that regime and for multiple observables are needed to maximize the scientific return of those surveys. Currently, one of the best ways of achieving this is by running cosmological hydrodynamic simulations. Unfortunately, those simulations are very computationally expensive, so only relatively small volumes of the Universe can be modelled. 

In this work, we explored the use of generative neural network models to make fast and accurate predictions of the distribution of one specific observable, the cosmic HI, over scales deep into the non-linear regime. Our WGAN model has identified a 100-dimension manifold, that characterizes the abundance and clustering of HI in the fully non-linear regime. By sampling from that manifold we produce new samples with very close statistical properties than the full hydrodynamic simulation IllustrisTNG. Moreover, we found that our generated samples exhibit better agreement with the IllustrisTNG simulated samples than the current benchmark HOD model when considering four summary statistics: the 1D PDF, the power spectrum, the bispectrum, and the void abundance. 

Our work demonstrates that GANs have the potential to learn how to accurately describe the 3D abundance and clustering properties of cosmic HI on scales as small as $35~h^{-1}{\rm kpc}$, and generate new samples by drawing from this underlying distribution. This allows us to generate new samples with the appropiate statistical properties five orders of magnitude more efficiently than IllustrisTNG. Our WGAN samples accurately reproduce the abundance of HI system across 9 orders of magnitude, beating state-of-the-art HOD models.

This works represents a first step towards a fast, accurate, and precise theory prediction pipeline needed to maximize the scientific return of upcoming cosmological 21cm survey missions. We are hoping that the variance of the produced samples can be made even closer to those of costly hydrodynamical simulations by conditioning the GANs predictions on the amplitude of the power spectrum in the predicted volume, an avenue which we plan to explore in follow-up work. In future work, we also plan to explore the possibility of creating samples of larger cosmological volumes and at different, lower redshifts. These, along with extensions of the model to predict HI velocities to allow a redshift-space analysis and extensions to different cosmological parameter values will, in our opinion, pave the way for machine learning methods to extend the capabilities of the small and expensive hydrodynamical simulations in the analysis of up coming survey data.

\section*{Acknowledgments}

We thank Gabriella Contardo and David Spergel for stimulating discussions. This project is supported by Center for Computational Astrophysics of the Flatiron Institute in New York City. The Flatiron Institute is supported by the Simons Foundation. This work was also supported in part through the NYU IT High Performance Computing resources, services, and staff expertise.

\section{Appendix}
\label{sec:appendix}

\subsection{WGAN \& MMDGAN in 2D}
\begin{figure*}
\begin{center}
  \includegraphics[width=0.7\linewidth]{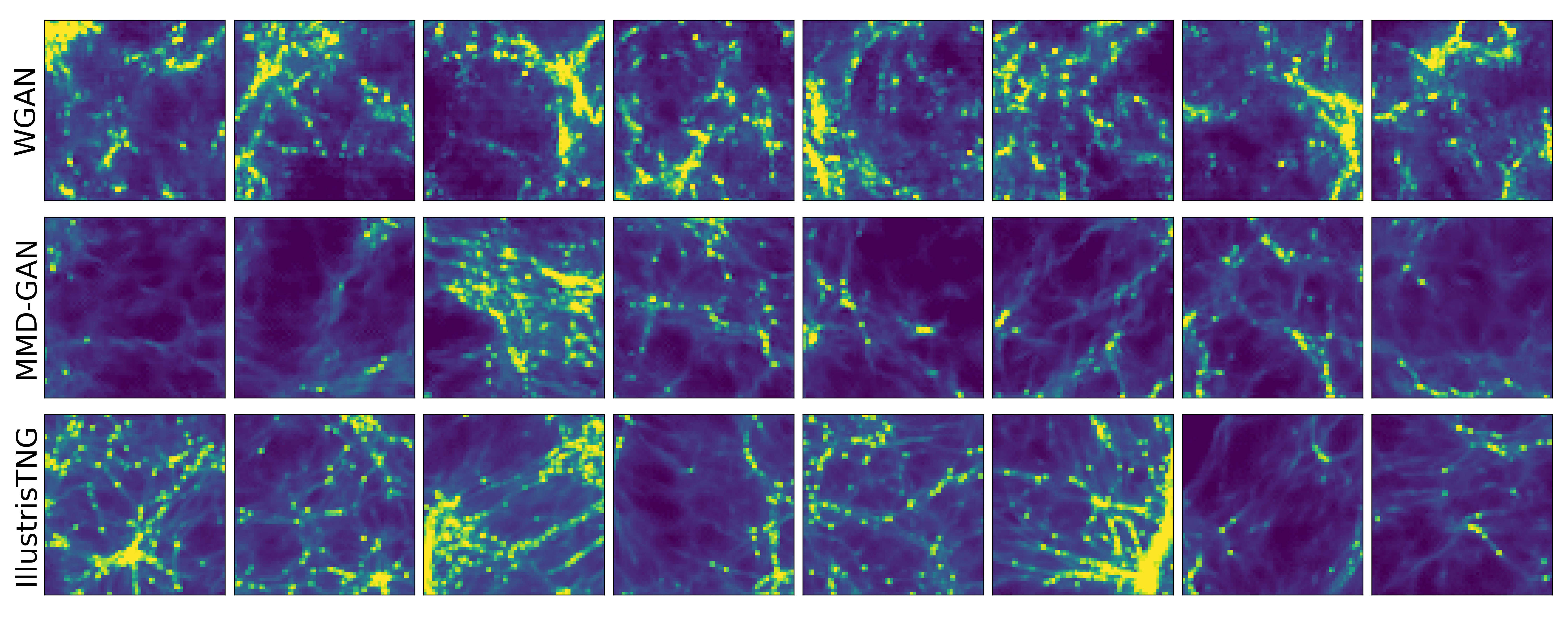}
  \caption{Visualizations of the 2D HI fields generated by the WGAN (top) and the MMD-GAN (middle). The MMD-GAN required significantly more hyperparameter tuning to reach convergence and results comparable in accuracy to those of the WGAN. The WGAN was trained for 45 epochs while the MMD-GAN was trained for 60 epochs. }
  \label{fig:2d_comparison}
  \end{center}
\end{figure*}

\begin{figure}
\begin{center}
    \includegraphics[width=.45\textwidth]{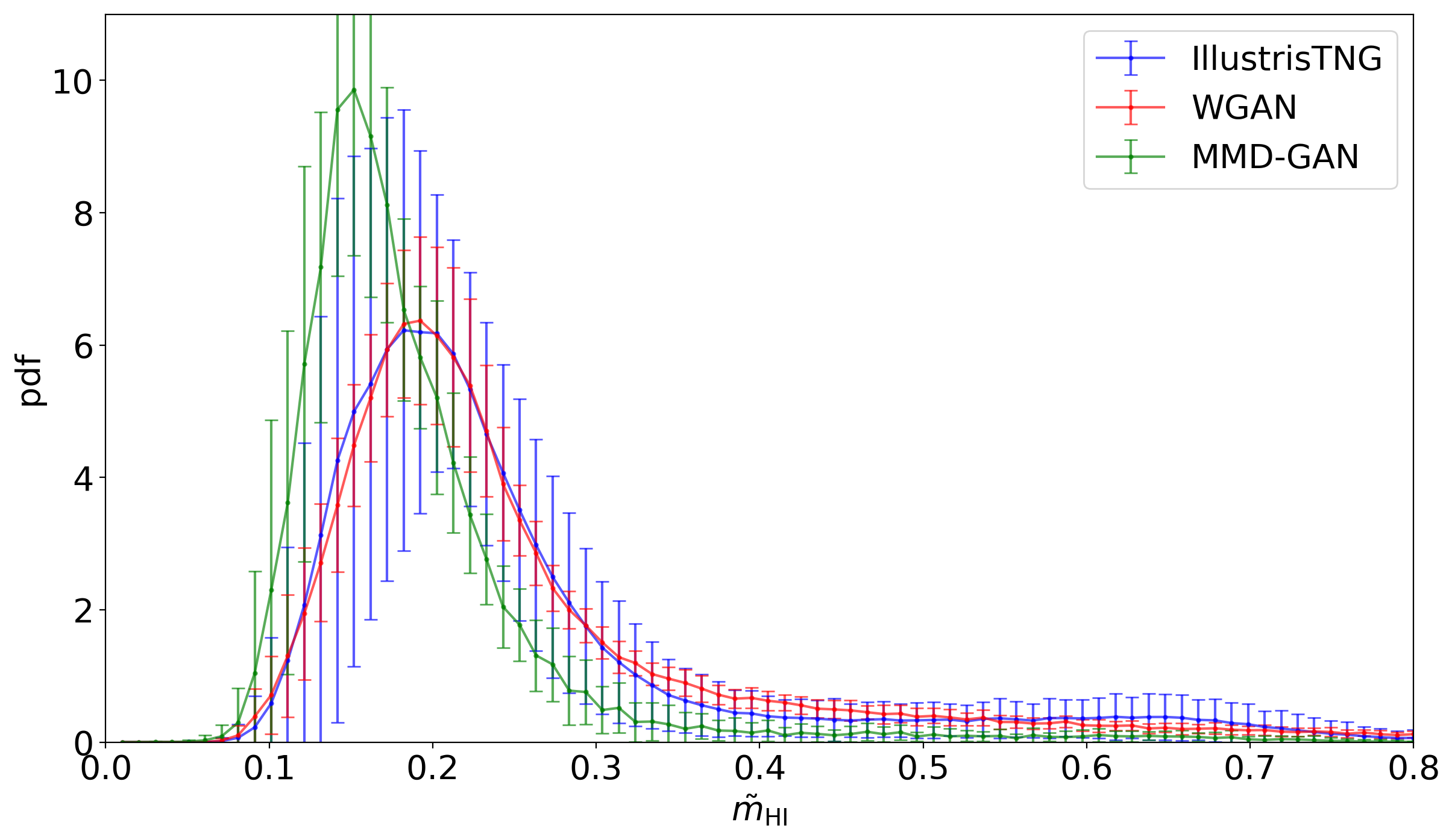}
\end{center}
\caption{Mean and the standard deviation of 1000 2D samples from IllustrisTNG (blue), WGAN (red) and MMD-GAN (green) of the 1D HI PDF as a function of HI mass. The WGAN samples match closely the HI 1D PDF of the IllustrisTNG samples when compared to the MMD-GAN samples.}
  \label{fig:2d_PDF}
\end{figure}

During this work, we explore the capabilities of two generative models for generating samples of HI distribution in 2D.
The choice of using WGAN for the generation of 3D samples of HI was influenced by the results of experiments we performed in 2D when comparing the two models. We observed that the WGAN model could produce good quality samples (see Fig. \ref{fig:2d_comparison}) significantly faster and with less need for hyperparameter tuning, whereas the MMD-GAN model needed a significant tuning of the parameters such as, but not limited to, kernel bandwidth, code space dimension, discriminator/generator training schedule and learning rates. In Fig. \ref{fig:2d_PDF}, we show a comparison between the 1D PDF from the 2D samples produced by the WGAN, MMD-GAN, and IllustrisTNG. It can be seen how the WGAN model outperforms the MMD-GAN in this case.

%
%
%
\bibliographystyle{lauapj}
\bibliography{refs}
%


\end{document}